\documentclass[twocolumn,showpacs,preprintnumbers,prl,floatfix, nofootinbib, superscriptaddress]{revtex4-1}
\usepackage{amssymb,amsmath, amsfonts}
\usepackage{graphicx,epsfig} 
\usepackage{bm} 
\usepackage{color}
\usepackage{soul}
\usepackage[breaklinks=true]{hyperref}

\definecolor{Myorange}{cmyk}{0,0.42,1,0}

\DeclareMathOperator{\Tr}{Tr} 

\begin{document}
\title{Detecting dynamic spatial correlation patterns\\
with generalized wavelet coherence and non-stationary surrogate data}
\author{M. Chavez}
\affiliation{CNRS UMR-7225, H\^{o}pital de la Piti\'e-Salp\^{e}tri\`{e}re. 75013 Paris, France}%

\author{B. Cazelles}
\affiliation{IRD-UPMC UMI-209, UMMISCO, 93143 Bondy, France}
\affiliation{CNRS UMR-8197, IBENS, Ecole Normale Sup\'erieure. 75005 Paris, France}

\begin{abstract}
Time series measured from real-world systems are generally noisy, complex and display statistical properties that evolve continuously over time. Here, we present a method that combines wavelet analysis and non-stationary surrogates to detect short-lived spatial coherent patterns from multivariate time-series. In contrast with standard methods, the surrogate data used here are realisations of a non-stationary stochastic process, preserving both the amplitude and time-frequency distributions of original data. We evaluate this framework on synthetic and real-world time series, and we show that it can provide useful insights into the time-resolved structure of spatially extended systems.  \end{abstract}
\pacs{02.50.Sk, 05.45.Tp, 05.45.Xt, 89.75.Fb}
\maketitle %

Synchronization is a fundamental phenomenon described in many biological and physical contexts for which there are two or more interacting oscillatory systems~\cite{synchEverything}. The interactions between coupled oscillators in real systems continuously create and destroy synchronised states, which can be observed as noisy and transient coherent  patterns.  The statistical detection of spatial synchrony in networks of coupled dynamical systems is therefore of great interest in disciplines such as geophysics, physiology and ecology~\cite{waveletsEverything, WaveletsAdvantages}. 

Statistical significance of transient coherent patterns cannot be assessed by classical spectral measures and tests, which require signals to be stationary~\cite{WaveletsAdvantages, WaveletsSignificance}. Synchrony estimators based on nonparametric methods have the advantage of not requiring any assumption on the time-scale structure of the observed signals. Among them, measures of synchrony or coherence based on wavelet transforms have been widely used to detect interactions between oscillatory components  in different real systems, i. e. neural oscillations, business cycles, climate variations or epidemics dynamics~\cite{waveletsEverything, WaveletsAdvantages}.

In recent years, different significance tests for the wavelet cross-spectrum or wavelet coherence have been developed to detect oscillatory patterns with covarying dynamics~\cite{WaveletsAdvantages, WaveletsSignificance, WaveletsSignificanceTESTS}. Unfortunately, the statistical assumptions of these tests are not always compatible with the structure of the data considered, and significance levels often depend on the structure of the wavelets applied~\cite{WaveletsSignificanceTESTS}. A rigorous theoretical framework cannot therefore be derived, and Monte Carlo simulations have to be performed to estimate the significance level~\cite{WaveletsSignificanceTESTS}. 

 \begin{figure*}[!htb]
   \centering
   \resizebox{0.95\textwidth}{!}{\includegraphics{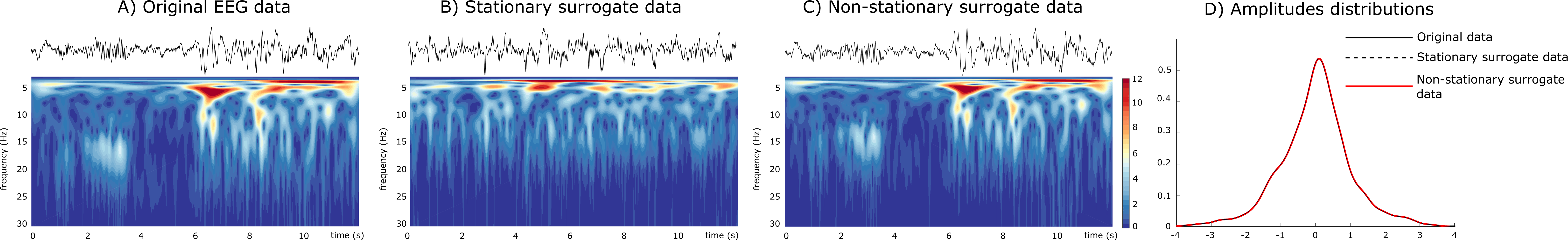}}
    \caption{Exemple of the time-frequency (TF) structure for different surrogate algorithms applied to epileptic EEG data. A) The original time series after reconstruction by the wavelet filtering, B) surrogate data generated with the  iterative Amplitude Adjusted Fourier Transform (iAAFT) algorithm, C) surrogate generated with our algorithm and D) distributions of amplitudes for the different algorithms. The color maps code for $\left | (W_x(t,f)) \right |$ values. Wavelet analysis of EEG were done over the frequency range $0-128$~Hz. For better visualisation spectra are displayed only for $f<30$~Hz.  Refer to Supplementary Material for a comparison with other algorithm (based on DWT) to replicate the TF structure.}  
   \label{figura1}
 \end{figure*}

Surrogate data techniques have been proposed as non-parametric resampling methods for testing general hypotheses on data without making assumptions on the underlying generating process~\cite{fftSurrogates}. However, time series measurements from real systems generally display irregular fluctuations, long-term trends, or a time-varying spectra. Such properties are incompatible with the main assumptions of standard surrogate data based on Fourier transform~\cite{FFT_AdvancedSurrogates, ARSurrogates}. 

Recently, parametric models have been also applied to test wider classes of null hypothesis, including non-stationary behaviour~\cite{ARSurrogates}.  Some limitations of these approaches include the relatively large basis dimension needed to obtain good optimisation, and the monitoring needed to control the  instabilities in the estimated model~\cite{TVARmodels}. Recent studies have proposed the use of discrete wavelet transforms  (DWT) for resampling time series such that the multiscale structure of original data is preserved~\cite{DWT_Surrogates, multifractalsSurrogates}. The main advantage of DWT is their ability to concentrate the signal's variance in a limited number of coefficients. Nevertheless, the number of data points heavily influences this decomposition (the number of scales); which may render the scale decompositions difficult to interpret~\cite{WaveletsAdvantages}. Although continuous wavelets often yield a redundant decomposition across scales, they are more robust to noise as compared with other decomposition schemes~\cite{WaveletsSignificance, WaveletsAdvantages, mallat1998}. 

In this work, we use a continuous wavelet-based approach to detect spatial coherent patterns in non-stationary multivariate observations. We generalise the wavelet coherence to multivariate time series and we extend the classic phase-randomised surrogate data algorithm to the time-frequency domain for generating non-stationary surrogates. This procedure preserves both the original amplitude and time-frequency energy (spectrogram) distributions. Compared with other surrogate algorithms, our method better replicate the time-frequency structure of real data. 
These non- stationary surrogates are used to assess the significance of transient coherent patterns found in multivariate time series. We evaluate the proposed method in different synthetic and real-world non-stationary data, and we show that this approach can substantially improve the detection of time-varying spatial coherence.  

We start by considering the time-frequency (TF) distributions obtained by correlating a time series $x(t)$ with a scaled and translated version of a chosen mother wavelet $w_{s,\tau}(t)=\frac{1}{\sqrt{s}}w(\frac{t-\tau}{s})$. In the following, we always consider the mother complex Morlet wavelet defined as $w(\theta) = \pi^{-1/4} \exp(-\theta^{2}/2) \times \exp(i \omega_0 \theta)$, where the parameter $\omega_0$ controls the time scale resolution of each wavelet. 

To quantify the relationships between two non-stationary signals, $x_i(t)$ and $x_j(t)$,  the wavelet cross-spectrum is given by $W_{i,j}(t,f)=W_{i}(t,f)W_{j}^*(t,f)$, where~$^*$ denotes the  complex conjugate operator and $W_{k}(t,f)$ is the wavelet transform of signal $x_k(t)$. Let us now consider $M$ zero-mean time series $x_1(t), \ldots, x_M(t)$, and define the complex coherence spectrum as $C_{i,j}(t,f)=\frac{\big \langle W_{i,j}(t,f) \big \rangle }{\left\| \big \langle W_{i,i}(t,f) \big \rangle \right \|^{\frac{1}{2}} \left \| \big \langle W_{j,j}(t,f) \big \rangle \right\|^{\frac{1}{2}}} $
for $i,j = 1, \ldots, M$, where $\langle\cdot\rangle$ denotes a smoothing operator both in time and frequency~\cite{waveletSmoothing}. 

 \begin{figure*}[!htb]
   \centering
   \resizebox{0.95\textwidth}{!}{\includegraphics{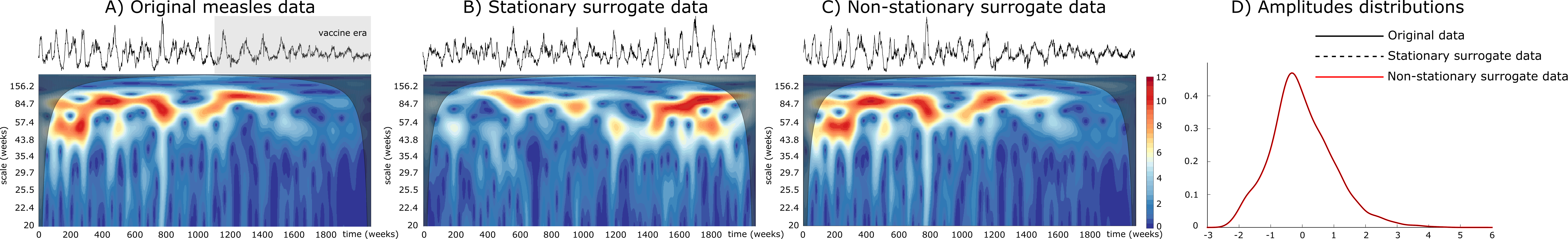}}
   \caption{Exemple of the TF structure for different surrogate algorithms applied to the squared root transformed measles data. Same stipulations as in the caption of Fig.~\ref{figura1}. Gray box in upper plot A indicates the vaccine era. Black transparent maps of time-frequency plots indicate the cone of influence that delimits the regions not influenced by edge effects~\cite{COI}.}  
   \label{figura2}
 \end{figure*} 
 
In bivariate data analysis, the wavelet coherence is defined as $\Gamma_{i,j}^2(t,f) = |C_{i,j}(t,f)|^2$. To extend this idea to the general case of $M\geqslant 2$ signals, we can define a matrix $\mathbf{\Sigma}(t,f)$ at every point in the time-frequency domain containing all the pairwise coherence spectra~\cite{likelihoodRatio}:
\begin{equation}\label{sigmaMatrix}
\mathbf{\Sigma}(t,f)=
\begin{bmatrix}
1 & C_{1,2}(t,f) & \ldots & C_{1,M}(t,f) \\
C_{2,1}(t,f) & 1 & \ldots & C_{2,M}(t,f) \\
\vdots & \vdots & \ddots  & \vdots \\
C_{M,1}(t,f) & C_{M,2}(t,f) & \ldots & 1  
\end{bmatrix},
\end{equation}

The time-varying spatial coherence (TVSC) can be defined by
\begin{equation}\label{TVSC}
\Psi(t,f) = \frac{1}{M-1}\left (\lambda_{\text{max}}^{\Sigma}(t,f)-1\right),
\end{equation}
where $\lambda_{\text{max}}^{\Sigma}(t,f)$ denotes the largest eigenvalue of the spectral matrix $\mathbf{\Sigma}(t,f)$.

The values of $\Psi(t,f)$ are bounded between $0\leqslant \Psi(t,f) \leqslant 1$, reaching the maximum when all the $M$ signals are locally -in the time-frequency plane- pairwise correlated ($\mathbf{\Sigma}(t,f)$ becomes an all-ones matrix with $\lambda_{\text{max}}^{\Sigma}(t,f)=M$); and the minimum when all signals are completely uncorrelated ($\mathbf{\Sigma}(t,f)= \mathbf{I}$ and $\lambda_{\text{max}}^{\Sigma}(t,f)=1$)\footnote{In case of stationary observations, eigenvalues of the covariance matrix are commonly used in radio communications for detecting spatial correlations between multivariate time series~\cite{likelihoodRatio}}. 

Interestingly, for the case $M=2$, $\mathbf{\Sigma}(t,f)$ is  given by the matrix $\begin{bmatrix}
1 & C_{1,2}(t,f) \\
C_{2,1}(t,f) & 1  
\end{bmatrix}
$, whose largest eigenvalue is $\lambda_{\text{max}}^{\Sigma}(t,f)=1 + |  C_{1,2}(t,f) |$, which yields $\Psi(t,f) = (\lambda_{\text{max}}^{\Sigma}(t,f)-1)= |C_{1,2}(t,f)|$. In the bivariate case, this therefore reduces the TVSC to the classic definition of the wavelet coherence $\Psi^2(t,f) = \Gamma^2(t,f)$. 

In wavelet-based analysis, test statistics are strongly affected by  data's structure, the mother wavelet's properties, and  by the smoothing applied~\cite{WaveletsSignificanceTESTS, waveletSmoothing}. In this work, the statistical properties of $\Psi(t,f)$ under the null hypothesis $H_0$ of $M$ uncorrelated processes are determined by Monte Carlo simulation. To do this, we generate a number of surrogate data realisations $\mathbf{\widehat{x}}^{j}(t), j=1,\ldots, K$ by repeating the randomisation procedure $K$ times. The statistical significance of $\Psi(t,f)$ values was assessed by a z-test to quantify the statistical deviation from values obtained in the ensemble of surrogate data. To correct for multiple testing, the false discovery rate (FDR) method was applied~\cite{FDR}. With this approach, the threshold of significance was set such that the expected fraction of false positives over the time-frequency plane is restricted to $q \leqslant 0.05$. 

A surrogate time series $\mathbf{\widehat{x}}(t)$ can be obtained by randomising the phase structure of the original signal $\mathbf{x}(t)$ in the time-frequency domain.  As the Morlet wavelet is a complex function, the resulting wavelet decomposition  $W_x(t,f)$ has both real and imaginary parts. We can therefore write $W_x(t,f)$ in terms of its phase $\phi_x(t,f) = \tan^{-1}\frac{\Im (W_x(t,f))}{\Re (W_x(t,f))}$ and modulus $\left | (W_x(t,f)) \right |$. The wavelet-based surrogate algorithm first generates a Gaussian white noise time series to match the original data length and it then derives the wavelet transform to extract the phase $\phi_\text{noise}(t,f)$. We use this randomised phase and the WT modulus of the original signal to obtain a surrogate time-frequency distribution $W_{\widehat{x}}(t,f) = \left | (W_x(t,f)) \right | \exp (i\phi_\text{noise}(t,f))$. A surrogate time series can be reconstructed by taking the real part of the inverse wavelet transform. Finally, the surrogate $\mathbf{\widehat{x}}(t)$ is rescaled to the distribution of the original data by sorting the data (after the wavelet filtering in the frequency band of interest) according to the ranking of the wavelet-based surrogate~\cite{fftSurrogates}. As its Fourier-based counterpart, our scheme can be iteratively repeated to better adjust the time-frequency distribution of the surrogate.

Other methods based on DWT have been proposed to resample non-stationary time series. Compared to standard surrogate data, they preserve better the TF structure of original data. Nevertheless, our algorithm replicates the TF distributions more accurately (see the Supplementary Material for additional details and comparisons).

To illustrate our surrogate data method, we consider an electroencephalographic (EEG) recording from a pediatric subject with intractable epileptic seizures~\footnote{The EEG data was obtained from the open repository CHB-MIT Scalp EEG Database (\url{https://www.physionet.org/pn6/chbmit})}. Although our approach is applicable to any neuroimaging functional method (e.g. EEG, fMRI, and MEG signals) here we use the EEG as this modality of acquisition has the major feature that collective neural behaviors, i.e., synchronization of cortical assemblies are reflected as time-varying interactions between EEG signals. The file studied here contains 22 EEG signals sampled at 256 Hz according to the 10-20 bipolar montage~\cite{EEGdataset}. 

The non-stationarity of  epileptic EEG signals is clearly illustrated in Fig.~\ref{figura1}-A. One can notice that the frequency content of epileptic oscillations may change rapidly across time over a range of frequencies. The time-frequency plot exhibits a short fast oscillatory behavior ($f \approx 15$~Hz) around $t=3$~s followed by slow and large oscillations accompanying the epileptic seizure after $t=6$~s. As depicted in Fig.~\ref{figura1}-B, classical stationary surrogate data (here we used the iAAFT algorithm~\cite{fftSurrogates}) is not able to replicate the non-stationary oscillations  embedded in the original signal. In contrast, the time-varying spectrum of original signal is clearly conserved with our algorithm, as illustrated in Fig.~\ref{figura1}-C. Plot in Fig.~\ref{figura1}-D confirms that the three surrogate algorithms conserve the amplitude distributions.

Another paradigmatic example of non-stationary  spatial synchrony is that observed in population dynamics. Here, we considered the weekly measles notifications in seven large English cities studied in Refs.~\cite{measles}.  Measles epidemics generally exhibit a non-stationary dynamics with a regular and highly epidemics before nationwide vaccination programs, and an irregular and spatially uncorrelated dynamics in the vaccine era. As illustrated in Fig.~\ref{figura2}, the data display multiannual cycles that dramatically varies with time, specially after vaccination. This rich behavior can not be encompassed by classical stationary surrogate data (Fig.~\ref{figura2}-B). Instead, the wavelet-based method perfectly keeps the variations of epidemic periods observed in the original time series (Fig.~\ref{figura2}-C). 

Throughout this work, the number of scales of the wavelet decomposition was selected such that an accurate reconstruction of the original signal was obtained, and the non-stationary oscillations and transient events observed in the time series were  accurately captured.

We then test now the performance of our framework to detect spatial coherent components on two synthetic datasets with time-varying  structure. In the first benchmark, the spatial system consists of 5 linear oscillators described by the following autoregressive (AR) model:
\begin{equation}\label{ARmodel}
\begin{split}
x_{1t} & = 0.95 \sqrt{2}x_{1(t-1)} - 0.9025x_{1(t-2)} +\epsilon_{1t}, \\
x_{2t} & = 0.6x_{2(t-1)}-0.3x_{2(t-2)}+k_2x_{1(t-1)}+ \epsilon_{2t}\\
x_{3t} &= 0.8x_{3(t-1)} - 0.5x_{2(t-2)} +0.4x_{1(t-1)}+\epsilon_{3t},\\
x_{4t} &=k_1x_{1(t-2)} +0.25\sqrt{2}x_{4(t-1)} + 0.25\sqrt{2}x_{5(t-1)} +\epsilon_{4t},\\
x_{5t} &= -0.25 \sqrt{2}x_{4(t-1)} + 0.25 \sqrt{2}x_{5(t-1)} +\epsilon_{5t}
\end{split}
\end{equation}
where $t$~denotes a discrete time index, $\epsilon_i$ are independent white noise processes with zero means and unit variances, and $k_i$ are coupling strengths. Here, we set $k_1=0$ and $k_2=0.15$ for $ t < 1000$; and $k_1=-0.5$ and $k_2=0.4$ for $ t \geqslant 1000$.  

Although the measure of time-varying spatial coherence is supposed to capture linear interactions, numerical evidence shows that $\Psi(t,f)$ still provides a qualitative description in case of nonlinear oscillators. Indeed, we consider a network of $i=1,\ldots,10$ coupled non-identical chaotic R\"{o}ssler oscillators. The equations of motion read
\begin{equation}\label{systemCoupledOscillators}
\begin{array}{lc}
\dot{x}_i = & -\omega_iy_i -z_i +\lambda \left[ \sum_j \xi{ij} (x_j-x_i)\right] +\sigma_i\eta_i,\\
\dot{y}_i =&\omega_{i}x_{i} + 0.165y_{i},\\
\dot{z}_i = &0.2 + z_{i}(x_{i}-10)
\end{array} 
\end{equation}
where $\lambda(t)$ is the time-varying coupling strength, $\xi{ij}$ are the elements of the coupling matrix (a random graph with an average number of links per node $k_m=4$); $\omega_i$ is the natural frequency of the $i^{\text{th}}$ oscillator (randomly assigned from a uniform distribution
with values between $0.98\leqslant \omega_i \leqslant 1.1$); $\eta_i$ denotes a Gaussian delta correlated noise with $\langle\eta_{i}(t)\rangle = 0$ and $\langle \eta_i(t)\eta_i(t') \rangle = 2 D \delta(t-t')$,  $D=0.01$. Coupling strength $\lambda$ varies with time as follows: $\lambda=0.5$ for $500 < t < 900$ and $\lambda=0.001$ elsewhere.
 
\begin{figure}[!htb]
   \centering
   \resizebox{\columnwidth}{!}{\includegraphics{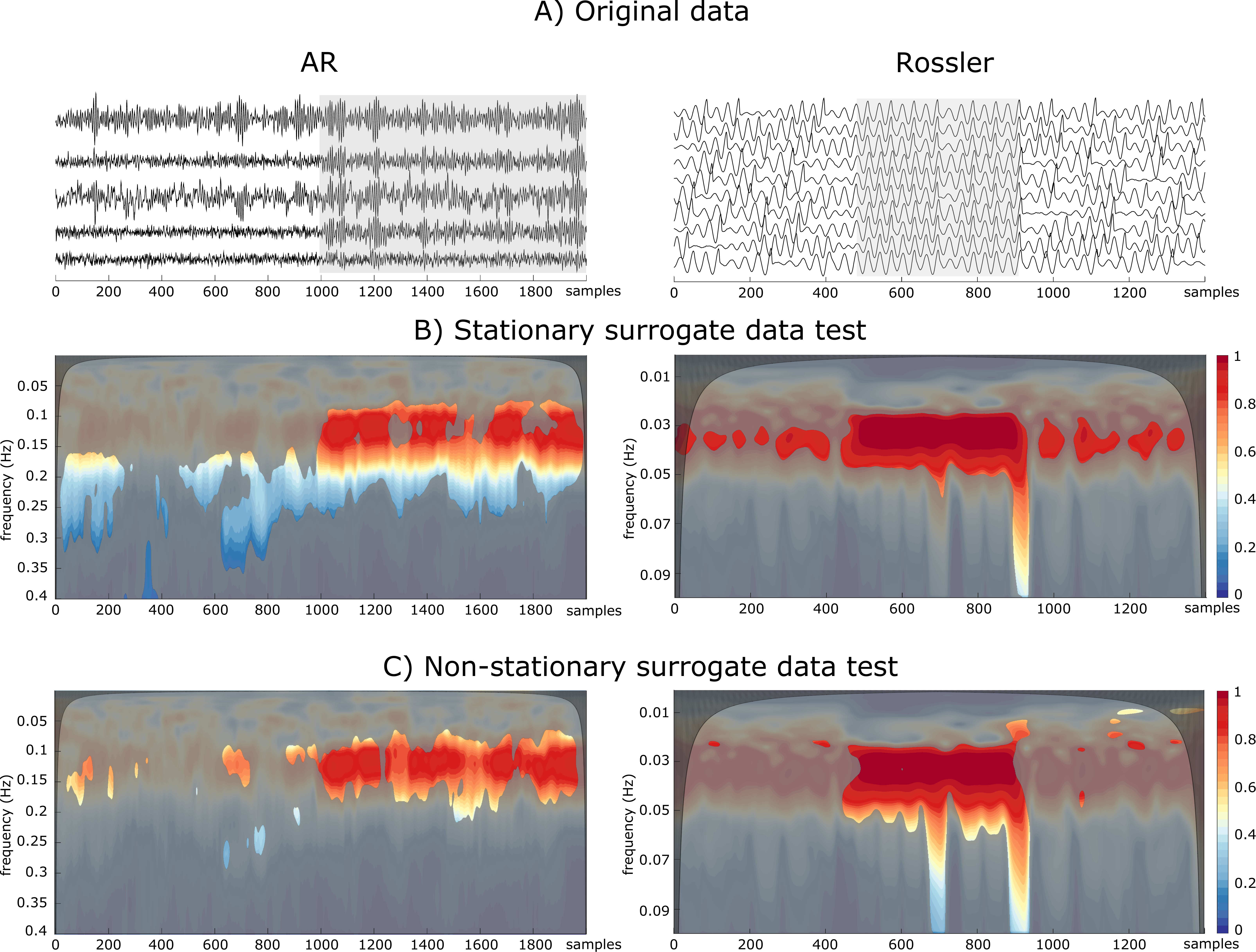}}
   \caption{$\Psi(t,f)$ values estimated from synthetic time series and statistical differences with those values obtained from different surrogate data. A) The original time series (gray boxes delimitate the region where the system is synchronized), B) surrogate data test based on the iAAFT algorithm and C) with our algorithm. The color maps code for $\Psi(t,f)$ values. Unmasked color regions in panels B-C indicate the significant levels. Black transparent maps indicate the cones of influence~\cite{COI}.}  
   \label{figura3}
 \end{figure}

Here, all time series are first centered and set to have zero mean and unit variance. Then, the TVSC values are computed. To estimate the distribution of  $\Psi(t,f)$ under $H_0$ we generate $100$ surrogates for each time series. 
 
In Figs.~\ref{figura3}-(A-C) we report, respectively, the time series generated by the above models, the significant coherent components detected by $\Psi(t,f)$ in combination with classical surrogate data and with the non-stationary surrogate data. Results reveal that stationary randomizations detect several large spurious synchrony patches on the time-frequency plane, e.g. the large patches before $t=1000$ for the coupled AR model, or those out of the synchronous region for the coupled R\"{o}ssler model ($500<t<900$). This is mainly due to the oscillations created over the whole segment by the stationary surrogate algorithm. Conversely, a detection based on our method considerably reduces the number of false coherent patches, while it clearly identifies the main regions with the highest spatial coherence. Remarkably, right plots of Figure~\ref{figura3} shows that the combination of $\Psi(t,f)$ with non-stationary surrogate data, constitutes a good criterion to assess spatial coherence in the case of nonlinear dynamical time series.

\begin{figure}[!htb]
   \centering
   \resizebox{\columnwidth}{!}{\includegraphics{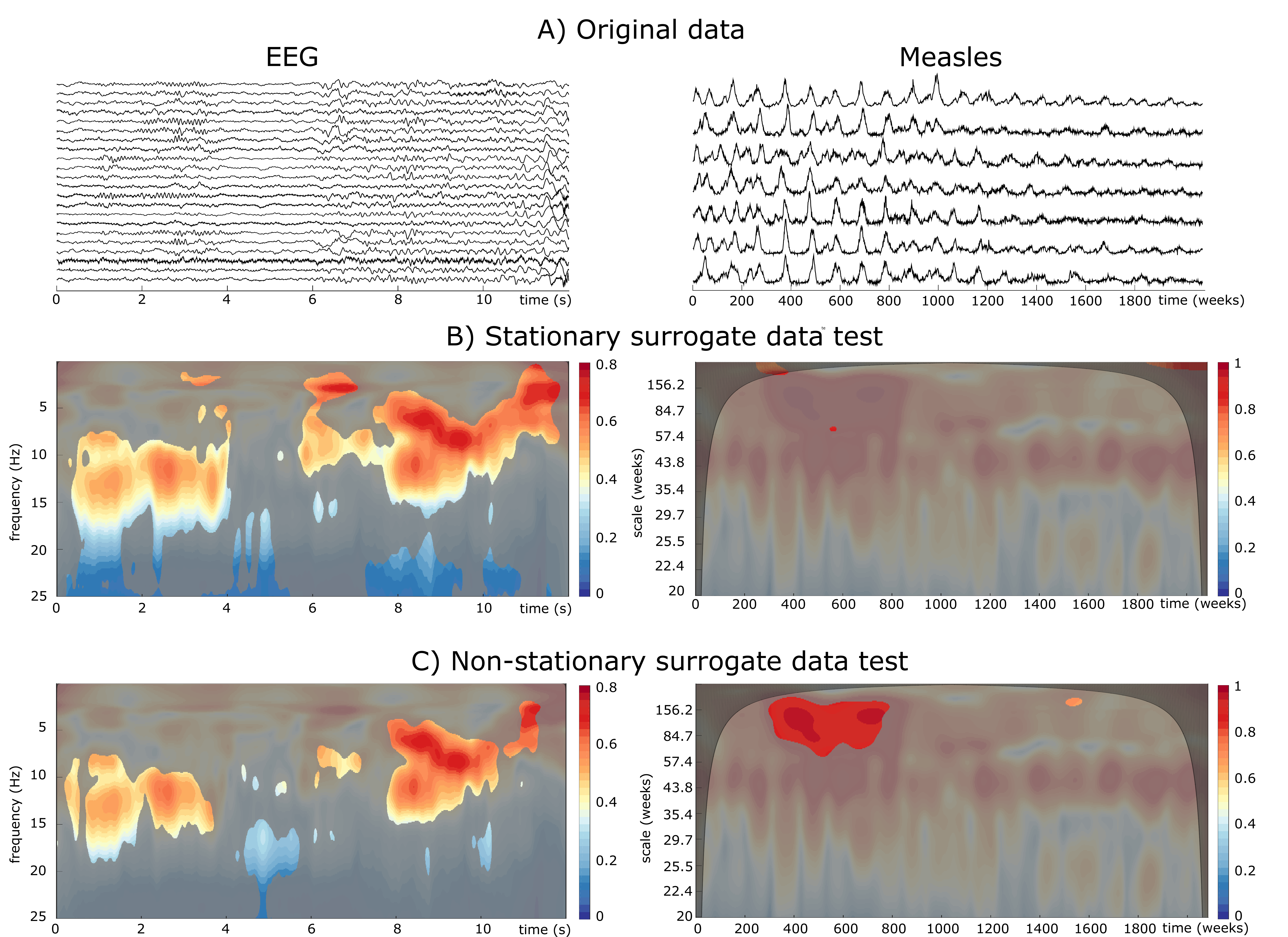}}
   \caption{$\Psi(t,f)$ values estimated from real spatial systems and statistical differences with those values obtained from different surrogate data. For measles data, missing values in each original time series were imputed using a local average, i.e. the mean of the two neighboring time points. Same stipulations as in the caption of Figure~\ref{figura3}. See Supplementary Material for a comparison with other algorithm (based on DWT) to detect these coherent patterns.}  
   \label{figura4}
 \end{figure}

To illustrate our approach on real-world time series, we study the two spatial systems described above. The situation with EEG data is illustrated in left panels of Figure~\ref{figura4}.  The first crucial observation is that, as expected in epilepsy dynamics, spatial coherent patterns are not time invariant, but instead they exhibit a rich time-frequency structure during seizure evolution. Results clearly show that classical surrogate data test may yield to the detection of large synchronous regions, specially at high frequency bands ($f \geqslant 20$)~Hz. In contrast, non-stationary surrogates improves the time-frequency localization of spatial correlation patterns. A first synchronous pattern seem to involve the low-amplitude fast oscillations often observed during the first seconds of epileptic seizures. Interestingly, the absence of significative values of $\Psi(t,f)$ between $t=4-8$~s suggest a desynchronization of some cerebral structures during the build-up of epileptic seizures, just before a wide synchronous spreading to the ensemble of the brain at $t=8$~s. This fully agrees with previous findings suggesting a neural desynchronization before the propagation of seizures which could facilitate the development of local pathological recruitments~\cite{neuralDesynch}.

Right panels of Figure~\ref{figura4} show the results for the measles data. We observe from $\Psi(t,f)$ values that global interactions between major epidemics change relatively smoothly through time. Classic surrogate analysis can capture epidemic's dynamics at different scales, but does not allow a proper description when they change with time. Indeed, standard surrogate data test reveals no significant spatial correlation patterns. Conversely, our approach clearly detects the main changes in spatial correlation structure: a high spatial coherence between the major epidemic (mainly biennial) component of time series is clearly identified in the pre-vaccine era. The interactions between the smaller epidemics with longer periods observed after vaccination are not found to be statistically significant. This is a remarkable result as it supports previous findings that during the pre-vaccination era, measles dynamics is characterized by a high spatial correlation of biennal epidemic patterns, while the vaccination eliminates large epidemics yielding thus a significant spatial decorrelation~\cite{measles}. 

In conclusion, we have addressed a fundamental problem in complex systems: detecting, from scalar observations, the time scales involved in spatial interactions of  oscillators with time-varying spectral components. Classical surrogate data tests require time-series to be stationary. Nevertheless, data recorded from real-world systems are generally noisy and non-stationary. In order to study their interactions we propose a complementary approach based on wavelet analysis. Wavelet coherence is generalized as a method for detecting transient but significant coherence between multivariate nonlinear signals. The classic surrogate algorithm is also generalized to produce non-stationary surrogates. Several artificial and real non-stationary, linear and nonlinear time series are examined, in order to demonstrate the advantages of our approach~\cite{suppCodes}.

Other wavelet-based methods have been used to analyse the relationships between multivariate signals~\cite{synchEverything, multivariateWavelets}. Nevertheless, standard significance tests assume stationarity of observations, which strongly affects the significance of the detected coherent patterns. Our results provide evidence of the constructive role of non-stationary surrogate data to uncover changes of correlation patterns in multivariate time series. Results confirm that our  method performs better than stationary Fourier-based and non-stationary DWT-based surrogate algorithms. The detection of spatial correlations in other multivariate data (e.g. financial or climate time series) might provide meaningful insights into the structure of other spatially extended systems.
\begin{acknowledgments}
BC is partially supported by the French Agence Nationale de la Recherche with the PANIC project (ANR-14-CE02-0015-01)
\end{acknowledgments}

\widetext
\begin{center}
\textbf{\large Supplementary Material}
\end{center}

\setcounter{equation}{0}
\setcounter{figure}{0}
\setcounter{page}{1}
\setcounter{footnote}{0}
\renewcommand{\theequation}{S\arabic{equation}}
\renewcommand{\thefigure}{S\arabic{figure}}
\renewcommand{\bibnumfmt}[1]{[S#1]}
\renewcommand{\citenumfont}[1]{S#1}
\renewcommand{\thefootnote}{S\arabic{footnote}}

\twocolumngrid

\noindent
\textbf{Surrogate data algorithms based on discrete wavelet transforms.}
Recent studies have propose the use of discrete wavelet transforms (DWT) for resampling time series such that the multiscale structure of original data is preserved. To this end, different randomization schemes have been used to resample the wavelet coefficients: random permutations, block resampling, or an iAAFT algorithm have been applied to wavelet coefficients to preserve the local mean and variance at each scale~\cite{S_breakspear2003, S_angelini2005, S_keylock}.  More refined permutations schemes have been proposed to preserve the interactions among scales reproducing thus the multifractal properties of the original data~\cite{S_palusSurrogates, S_keylockMultiFractal}. 

In the surrogate algorithm proposed by Ref.~\cite{S_palusSurrogates} wavelet coefficients are resampled (permuted) on a dyadic tree that, when inverted, yields a data set with  multifractal characteristics as the original data. Although an amplitude adjustment is used to recover the original distribution of wavelet coefficients at each scale of the wavelet decomposition, the histogram of the new time series does not exactly replicate the histogram of the original data. A more refined method (the iterative Amplitude Adjusted Wavelet Transform, iAAWT) is proposed in Ref.~\cite{S_keylockMultiFractal} in which the wavelet decomposition is obtained by means of a dual-tree complex DWT. At each scale of the decomposition, phases are extracted and randomized. These phases are then combined with the original amplitudes and the inverse wavelet transform is applied to obtain a new time series.  An amplitude adjustment is applied to preserve the histogram of the original data. Last steps  are iterated until convergence. Compared with other methods based on DWT, this algorithm has been shown to be more accurate and precise to approximate the multiscale structure of time series with a time-varying spectrum.

The main advantage of all the methods based on a DWT is their ability to concentrate the signal's variance in a limited number of scales. Nevertheless, as these decompositions are generally computed on a dyadic base (the number of scales is always $2^n$), the number of data points heavily influences these decompositions and particularly the time-scale localization of the signal's variance; which may render the scale decompositions difficult to interpret, especially their time evolution~\cite{S_WaveletsAdvantages}.

\bigskip
\noindent
\textbf{Replication of the time-frequency structure with other algorithm.} To compare the ability of our method to replicate the time-frequency (TF) distribution of the original data, we estimate the error:
\begin{equation}
e^2 =  \frac{\left\| \left | W_{\widehat{x}}(t,f) \right | - \left | W_{x}(t,f) \right | \right\| _{F}}{\left\| \left | W_{x}(t,f) \right | \right\| _{F}}
\end{equation}
were  $\left\| A \right\| _{F}=\sqrt{\Tr (A A^{T})}$  denotes the Frobenius norm of real matrix $A$, and $W_{\widehat{x}}(t,f)$ and $W_{x}(t,f) $ are the wavelet transforms of surrogate and original data, respectively.

\bigskip
 \begin{figure}[!ht]
   \centering
   \resizebox{0.9\columnwidth}{!}{\includegraphics{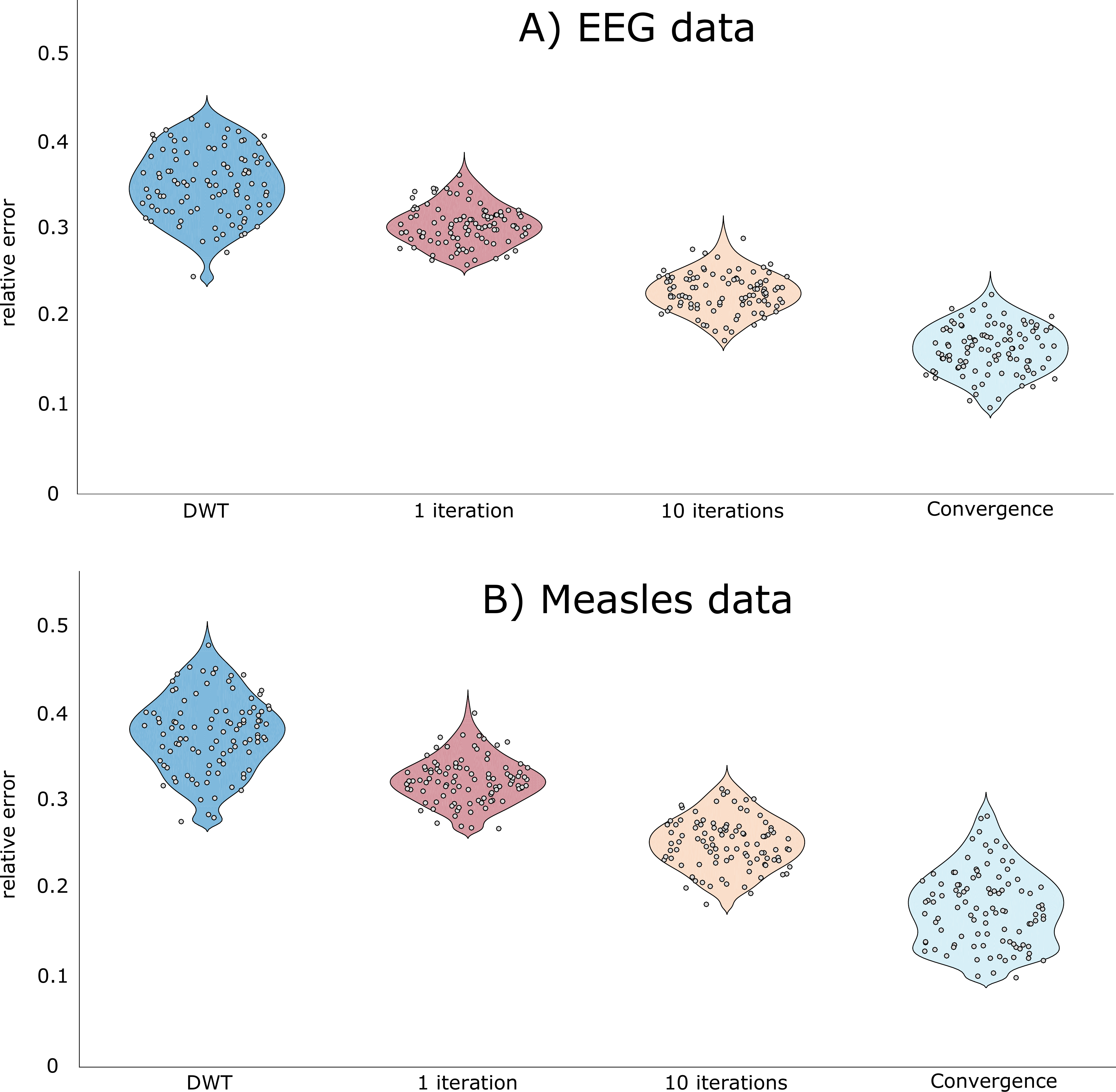}}
    \caption{Comparsion to preserve the TF distributions of the EEG and measles data. Violin plots represent distributions of error values over 100 surrogates (gray points). The error of iAAWT algorithm is significantly larger than those obtained by our method at different number of iterations (a right-tailed Wilcoxon rank-sum test with alpha set at $p = 0.01$).}  
   \label{figure1SuppMat}
 \end{figure}

 \begin{figure*}[!htb]
   \centering
   \resizebox{0.95\textwidth}{!}{\includegraphics{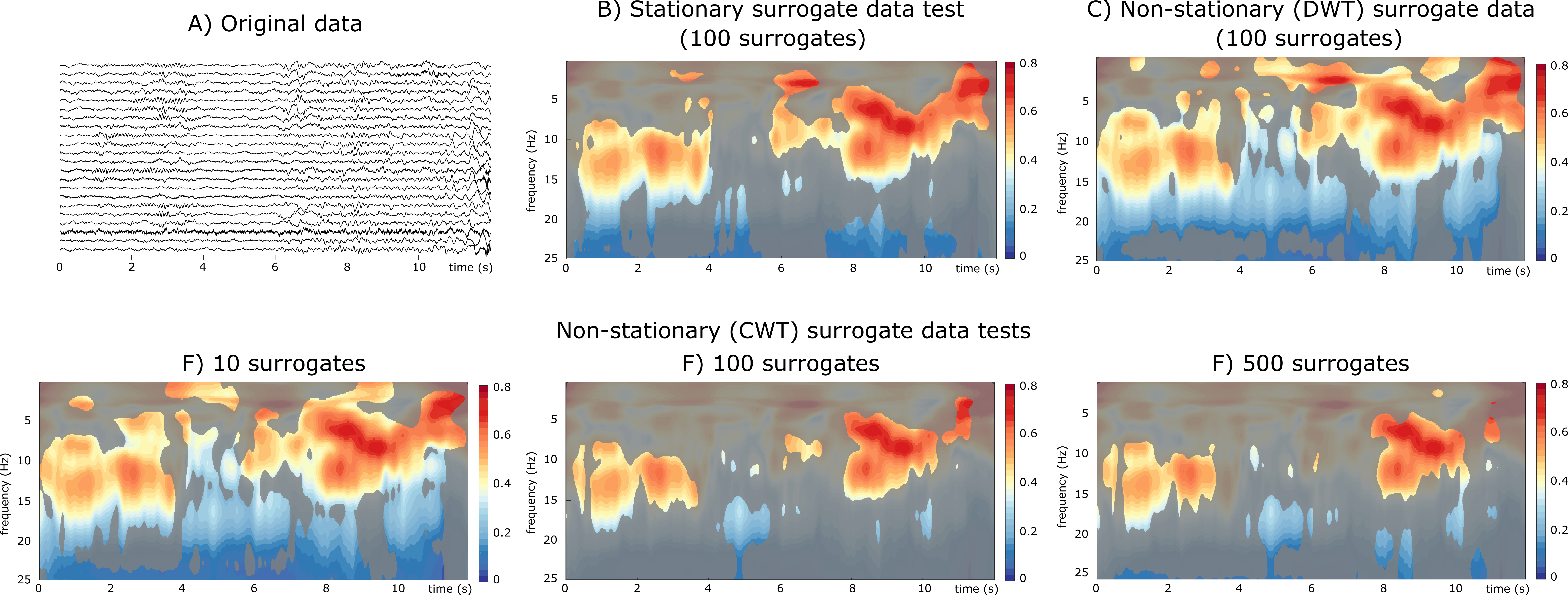}}
    \caption{$\Psi(t,f)$ values estimated from epileptic EEG data and statistical significances obtained from different surrogate data tests. A) The original time series; B) surrogate data test of the time-varying spatial coherence with the iterative Amplitude Adjusted Fourier Transform (iAAFT) algorithm; C) non-stationary surrogate data test with the iAAWT algorithm; and significance tests  D) $N=10$, E) $N=100$, and F) $N=500$ surrogate data obtained with the continuous wavelet transform (CWT). Color maps code for $\Psi(t,f)$ values. Unmasked color regions in panels B-F indicate the significant levels. Black transparent maps indicate the cones of influence that delimit the time-frequency regions not influenced by edge effects.}  
   \label{figure2SuppMat}
 \end{figure*}

To compare our method, different surrogating algorithms are applied to  the two time series studied in Figs.~1 and 2 of the main text (the EEG and measles time series).  We firstly compare our algorithm with the iAAWT method (refer to Ref.~\cite{S_keylockMultiFractal} for detailed description of the algorithm) \footnote{We have used the code available at the File Exchange site of MathWorks \href{https://fr.mathworks.com/matlabcentral/fileexchange/62382-iterated--amplitude-adjusted-wavelet-transform--iaawt--for-time-series-randomisation?}{\textbf{here}}}. We also apply an improved iterative version of our method, in which an approximation of the original TF distribution is refined with each iteration until it is determined to be sufficiently accurate, at which time the procedure terminates. 

In Fig.~\ref{figure1SuppMat} we report the normalized error produced by different surrogate replicates of original TF distributions. By construction, the histograms of the original data are exactly replicated by all the algorithms. For each time series, 100 surrogates are generated and the error computed. Our algorithm is applied in its simple version (a single iteration), after ten iterations, and finally for more than 30 iterations. Although the iAAWT algorithm replicates the multiscale structure of original data, our algorithm does a better job in this respect. Even after a single iteration, our algorithm reproduces better the non-stationary oscillations of studied time series. The error differences between the iAAWT and our method are all significant as assessed by a non-parametric test, the Wilcoxon-Mann-Whitney test, with alpha set at $p = 0.05$.

\noindent
\textbf{Detection of coherent patterns with different surrogates. --} 
We study more in detail the performances of our algorithm to detect non-stationary spatial coherent components. In addition to the standard stationary surrogate data test, we also apply the iAAWT to the same real-world multivariate data studied in the main text: the epileptic EEG recordings and the measles dataset. We also evaluate the  detection of significant coherent patterns for different number of surrogates generated by the simplest version of our algorithm (a single iteration). Significant coherent patterns were detected as statistically different from those obtained from surrogates, by a z-test corrected by a FDR at $q \leqslant 0.05$.

Results in Fig.~\ref{figure2SuppMat} clearly show that, for the EEG data, a test of significance based on the non-stationary iAAWT method yields to the detection of large and spurious synchronous regions, as those detected between $15-25$~Hz during practically the whole recording (these fast oscillations are concentrated mainly around $0< t < 4$). Similar results are obtained with a reduced number of surrogates (10 realisations). In contrast, for 100 or 500 realisations, the detection based on our method considerably reduces the number of false coherent patches, and it clearly identifies the main regions with the highest spatial coherence. 

\begin{figure*}[!htb]
   \centering
   \resizebox{0.95\textwidth}{!}{\includegraphics{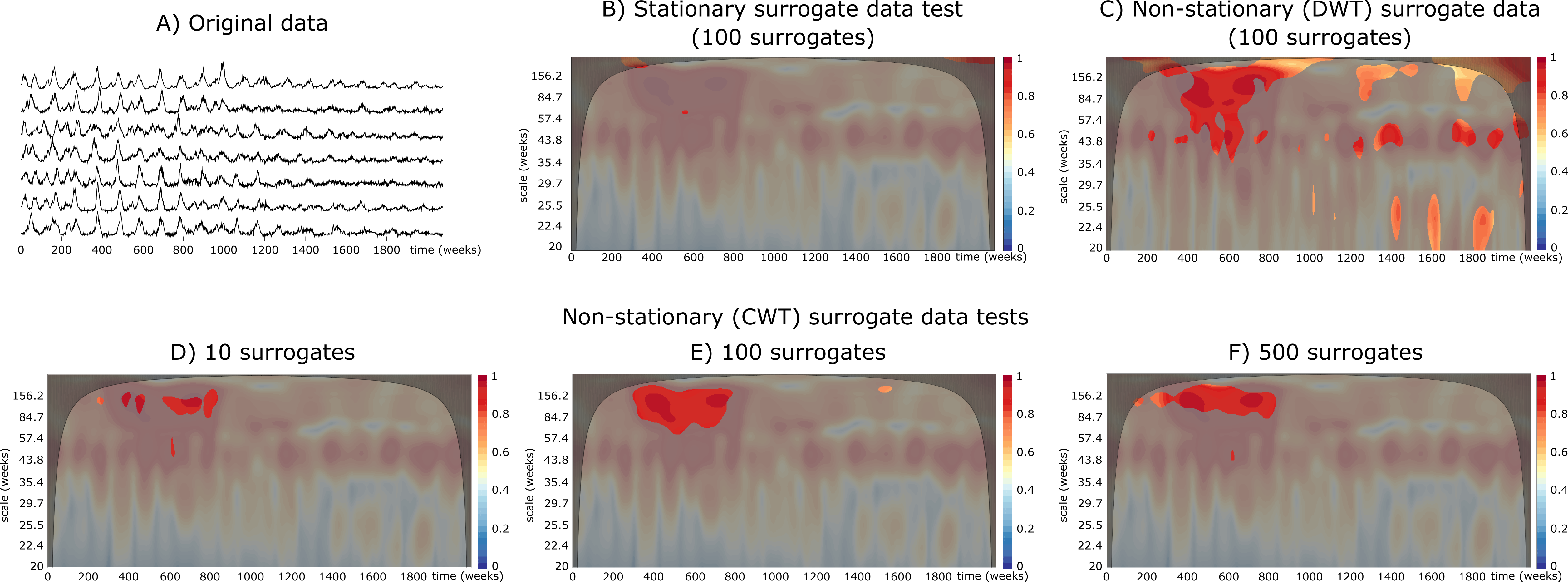}}
    \caption{$\Psi(t,f)$ values estimated from measles data and statistical significances obtained from different surrogate data tests. Missing values in each original time series were imputed using a local average, i.e. the mean of the two neighboring time points. Same stipulations as in the caption of Figure~\ref{figure2SuppMat}}  
   \label{figure3SuppMat}
 \end{figure*}

Results shown in Figure~\ref{figure3SuppMat} indicate that, for the detection of coherent spatial patterns in measles data, tests based non-stationary surrogate perform better than standard surrogate data test. The use of surrogates obtained by the iAAWT  algorithm reveals the major biennal synchronous epidemic component characteristic of the pre-vaccine era. Nevertheless, it also detects short periods of spatial interactions between annual oscillations, often associated to spurious correlations produced by seasonal variations~\cite{S_spuriousCorrelations}. For a reduced number of non-stationary surrogate time series, practically no significant spatial coherent patterns are detected. In contrast , when the number of surrogate is increased (100 or 500) our approach clearly identifies the high spatial correlation between the major epidemic (mainly biennial) component of time series in the pre-vaccine era. 

These supplementary results suggest that, when a sufficient number of surrogates is applied, our test constitutes a good criterion to assess spatial coherence in the case of time series with time varying spectra. Applied to other non-stationary time series, these surrogates could also be used to assess the statistical significance of different time-varying statistics, such as time-varying causality, evolutionary bispectrum, dynamic phase or cross-frequency interactions, among others~\cite{S_timeVaryingNonlinearities}.


\begin{thebibliography}{99}
\bibitem{COI}
As the wavelet is centered close to the edges of time series, edge effects occur. The area of the TF plane where such effects are relevant, the so-called cone of influence, was chosen as the $e$-folding time of the Morlet wavelet function~\cite{WaveletsSignificance}.
\end{thebibliography}
\end{document}